# Modal Analysis Investigation of Mechanical Kerr Frequency Combs

**Samer Houri[1], Daiki Hatanaka[1], Yaroslav M. Blanter[2], Hiroshi Yamaguchi[1]**

[1] NTT Basic Research Laboratories, NTT Corporation, 3-1 Morinosato-Wakamiya, Atsugi-shi, Kanagawa 243-0198, Japan.

Houri.Samer@lab.ntt.co.jp, samer.houri.dg@hco.ntt.co.jp

[2] Kavli Institute of Nanoscience, Delft University of Technology, Lorentzweg 1, 2628 CJ Delft, The Netherlands.

**Abstract.** The aim of this work is to theoretically investigate the possibility of Kerr frequency combs in mechanical systems. In particular, whether microelectromechanical devices (MEMS) can be used to generate frequency combs in a manner that is analogous to the optical frequency combs generated in optical microresonators with Kerr-type nonlinearity. The analysis assumes a beam-like micromechanical structure, and starting from the Euler-Bernoulli beam equation derives the necessary conditions in parameter space for the comb generation. The chapter equally presents potential means for the physical implementation of mechanical Kerr combs.

## 1    Introduction

Optical frequency combs, as their name suggests, produce an output that has equidistant frequency components within a relatively wide spectral range. The time domain corollary of this discrete and equidistant multispectral structure is that the output takes the form of a periodic train of narrow pulses (Hall 2005). The highly periodic signal output and equidistant spectral components have made optical frequency combs an invaluable tool for timing and spectroscopy applications (Eckstein et al. 1978, Udem et al. 1999, Diddams 2000, Ma et al. 2004).

Several means exist for the generation of optical frequency combs, in one approach mode-locked cavities such as Ti: Sapphire laser cavities that can support a large number of longitudinal modes is used to provide the equidistant components of the optical comb (Spence 1991, Cundiff et al. 2001). In a slightly different implementation, known as the Q-switched cavity, a phase relation is established between the different cavity modes using a saturable absorber (Haus 1976).

More recently a very promising approach for the generation of optical frequency combs making use of Kerr nonlinear optical resonators has been demonstrated. The basic principle behind the operation of such Kerr frequency combs consists of applying a laser pump that is nearly resonant with one of the cavity's many modes, the laser pump is thus confined to micrometric dimensions inside the resonator resulting in very large electric fields. The intra-cavity fields are large enough to undergo nonlinear degenerate four-wave mixing (FWM) which produces photons with slightly different frequencies that excite nearby cavity modes. These newly excited modes in their turn undergo a process of either degenerate or non-degenerate FWM to produce more sidebands that correspond to other cavity modes, and so on in a cascade process (Kippenberg et al. 2004, Del'Haye et al. 2007).

The introduction of Kerr frequency combs has sparked great interest that is motivated by fundamental as well as practical questions. On a more technological end a significant appeal of Kerr frequency combs is their use of microscale optical resonators thus promising to reduce both footprint and power consumption considerably compared to the more classical mode-locked lasers approach (Levy et al. 2010, Kippenberg et al. 2011, Stern et al. 2018, Spencer et al. 2018). Whereas the generation of solitons (Grelu and Akhmediev 2012, Herr et al. 2013, Marin-Palomo et al. 2017, Bao et al. 2017, Obrzud et al. 2017), rogue waves (Coillet et a. 2014, Akhmediev et al. 2013), emerging patterns (Lugiato and Lefever 1987), and chaotic behavior (Matsko et al. 2013, Coillet and Chembo 2014, Panajotov et al. 2017) are all issues of importance from a more fundamental point of view.

Seen the great progress and interest in photonic frequency combs, it is not surprising that phononic, i.e. mechanical or acoustic, frequency combs should also be investigated. This is especially true since micromechanical systems (MEMS) offer strongly nonlinear behavior (Lifshitz and Cross 2008) that is promising for frequency comb implementation. Indeed several approaches have been undertaken to demonstrate frequency combs, the first approach relies on mixing two drive tones using the nonlinearity of the MEMS device (Erbe et al. 2000, Jaber et al. 2016, Hatanaka et al. 2017, Houri et al. 2019). A second approach relies on inducing instabilities in a highly nonlinear M/NEMS device, such instabilities usually originates from the nonlinear interaction between different oscillating modes or devices (Karabalin 2009, Mahboob et al. 2012, Cao et al. 2014, Mahboob et al. 2016, Houri et al. 2017, Seitner et al. 2017, Ganesan et al. 2017, Ganesan et al. 2018, Houri et al. 2018, Czaplewski et al. 2018).

Despite such progress in demonstrating mechanical frequency combs, these demonstrations remain limited to combs generated within the envelope of a single nonlinearly oscillating mode, while a true multimodal mechanical frequency comb with a large number of cavity modes excited simultaneously, in a manner analogous to optical frequency combs remains to be demonstrated. Indeed even the questions regarding the possibility and properties of a Kerr mechanical frequency comb have not been addressed in a satisfactorily fashion.

Therefore, it is the aim of this work to explore the possibility of producing a multimodal mechanical frequency comb that functions according to the same principals of that of optical Kerr combs. The theoretical investigation will further explore the conditions in parameter space that are necessary for the onset of such combs, in the aim of guiding future experimental effort. Thereafter, the chapter proposes experimental means for the generation of mechanical combs.

## 2    Analytical Formulation

In addition to numerous experimental investigations, optical Kerr combs have equally benefited from a detailed and extensive mathematical analysis of their properties, formation conditions, and dynamics (Matsko et al. 2005, Chembo and Menyuk 2013, Godey et al. 2014), whereas such analytical basis is still lacking for the case of mechanical Kerr combs.

This section will provide mathematical grounding that establishes a relation between beam mechanics and the dynamics of FWM Kerr comb generation. Once a mechanical analogue to the optical governing equations is established, attention is given to the physical interpretation of these equations and their corresponding parameter space. This section will equally investigate the impact of some MEMS-specific properties such as internal stresses that are often encountered in MEMS devices but have no analogue in optics. This work assumes a beam-like geometry that is compatible with the Euler-Bernoulli beam equation.

### 2.1   Euler-Bernoulli modal expansion

We start our investigation by considering the beam equation to describe the dynamics of MEMS beam structures. We then decompose the mechanical vibrations of the beam into a set of normal modes in a manner that is analogous to the modal decomposition of optical microresonators (Chembo and Yu 2010). The beam equation used in this work is the Euler-Bernoulli equation which along with the boundary conditions for a clamped-clamped beam, reads (Cleland 2003):

$$\begin{cases} EI\frac{\partial^4 y}{\partial x^4} + \gamma\frac{\partial y}{\partial t} + \rho A\frac{\partial^2 y}{\partial t^2} - \frac{EA}{2L}\int_0^L \left(\frac{\partial y}{\partial x'}\right)^2 dx' \frac{\partial^2 y}{\partial x^2} = F(x,t) \\ \qquad\qquad y(0,t) = y(L,t) = y'(0,t) = y'(L,t) \end{cases} \qquad (1)$$

where $x$ and $y(x,t)$ are the longitudinal coordinate and the beam's displacement respectively, $E$ is the Young's modulus, $I$ is the second moment of inertia, $\gamma$ is the viscous damping term, $\rho$ is the density of the beam, $A$ is the beam's cross-sectional area, $L$ is the beam's length, $F(x,t)$ is the driving force applied to the structure, $x'$ is also a longitudinal coordinate used only within the integral and the last term on the left hand side is the nonlinear term originating from beam stretching (Cleland 2003).

In order to make this description scale free, the above equation is changed into a dimensionless version by resorting to the following set of parameter transformation:

$$\bar{x} = \frac{x}{L}, \bar{y} = \frac{y}{h}, \bar{\gamma} = \frac{\gamma L^4}{EIT}, \bar{F} = \frac{FL^4}{EIh}, \bar{t} = \frac{t}{T}, T = \sqrt{\frac{\rho AL^4}{EI}} \qquad (2)$$

where the bar indicates a dimensionless quantity.

The dimensionless Euler-Bernoulli equation now reads (dropping the bars for convenience):

$$\frac{\partial^4 y}{\partial x^4} + \gamma \frac{\partial y}{\partial t} + \frac{\partial^2 y}{\partial t^2} - 6 \int_0^1 \left(\frac{\partial y}{\partial x'}\right)^2 dx' \frac{\partial^2 y}{\partial x^2} = F(x,t) \qquad (3)$$

At this point we resort to a modal decomposition where the motion of the vibrating beam is projected on a set of harmonic basis functions, i.e. $y(x,t) = \sum_i \psi_i(x)\xi_i(t)$. These latter are simply the natural modes of vibration of a linear beam, and can equally be obtained as the homogenous solution of the Euler-Bernoulli equation, written as (the same boundary conditions apply):

$$\frac{d^4 \psi(x)}{dx^4} = \omega^2 \psi(x) \qquad (4)$$

Whose solution i.e. mode shapes for a clamped-clamped beam are given as:

$$\psi_n(x) = a_n\left(\cos(\omega_n^{1/2} x) - \cosh(\omega_n^{1/2} x)\right) + b_n\left(\sin(\omega_n^{1/2} x) - \sinh(\omega_n^{1/2} x)\right) \qquad (5)$$

And the normal mode frequencies are obtained from numerically solving the following transcendental function:

$$\cos(\omega_n^{1/2} x)\cosh(\omega_n^{1/2} x) = 1. \qquad (6)$$

Inserting the harmonic summation into the partial differential equation (3), the latter now reads:

$$\sum_k \left(\xi_k \frac{d^4 \psi}{dx^4} + \gamma_k \psi_k \dot{\xi}_k + \psi_k \ddot{\xi}_k - 6 \int_0^1 (\sum_l \sum_m \psi_l' \psi_m') dx' \psi_k'' \xi_k \xi_l \xi_m\right) = F(x,t) \qquad (7)$$

where $(\bullet)$ indicates time derivative, and $(')$ indicates spatial derivative. The orthonormality of the mode shapes imply the following identities:

$$\int_0^1 \psi_n \psi_m \, dx' = 0; \ \int_0^1 \psi_n \psi_n \, dx' = 1. \tag{8}$$

Multiplying equation (7) by $\psi_n(x)$ and integrating from 0 to 1, we obtain the following modal equation:

$$\ddot{\xi}_n + \gamma_n \dot{\xi}_n + \omega_n^2 \xi_n + 6 \sum_k \sum_l \sum_m \int_0^1 (\psi_k' \psi_l') dx' \int_0^1 (\psi_m' \psi_n') dx' \, \xi_k \xi_l \xi_m = F_n(t) \tag{9}$$

where $\gamma_n$ is the modal damping, $F_n$ is the modal forcing given as $F_n(t) = \int_0^1 F(x,t) \psi_n \, dx$, and we have modified the nonlinear term using the identity $\int_0^1 (\psi_n' \psi_n') dx' = -\int_0^1 (\psi_m \psi_n'') dx'$.

We define the following equality, $\Lambda_{mn} = \int_0^1 (\psi_m' \psi_n') dx'$, thus using these identities the partial differential equation (3) can finally be written as the following set of ordinary differential equations:

$$\ddot{\xi}_n + \gamma_n \dot{\xi}_n + \omega_n^2 \xi_n + 6 \sum_{klm} \Lambda_{mn} \Lambda_{kl} \xi_k \xi_l \xi_m = F_n(t) \tag{10}$$

Note that the obtained equation, reduces to a classical Duffing equation if the number of modes are reduced to one, and reduces to a mode-coupling equation (Westra et al. 2010, Lulla et al. 2012, Matheny et al. 2013, Yamaguchi et al. 2013) if only two modes are allowed.

To proceed, a rotating frame approximation (Greywall et al. 1994) is applied whereby the motion of the structure's modes is expressed as:

$$\xi_j = \frac{1}{2} \big( A_j e^{i\omega_j t} + A_j^* e^{-i\omega_j t} \big); \text{ and } F_n(t) = \frac{F_n}{2} (e^{i\omega t} + e^{-i\omega t}) \tag{11}$$

where $A_j$ is a slowly varying complex envelope that will capture the main dynamics of the $j$th mode, $F_n$ and $\omega$ are the amplitude and frequency of the driving force respectively, with $\omega$ being very close to $\omega_n$.

Note that since we define the amplitude envelope to be slowly varying, that implies that our rotating frame can only account for effects that take place on time scales that are $\gg 1/\omega$, a subtlety that will have important implications later on.

Furthermore, we define $\omega = \omega_n(1 + \delta)$, inject equation (11) into equation (10), and drop all second order terms of $\delta$, e.g. $\delta^2$ and $\delta\gamma$, keeping only the first order time derivative of the complex amplitude envelope, and removing all frequency terms that are not on the order $\sim \omega$ (i.e. dropping all terms that are on the order $2\omega$, $3\omega$ …) we obtain the following:

$$i\omega_n \dot{A}_n - \delta \omega_n^2 A_n + \frac{i}{2} \gamma_n \omega_n A_n + \frac{3}{4} \sum_{k,l,m} \Lambda_{kl} \Lambda_{mn} A_k A_l^* A_m e^{-i\Delta\omega_{klmn} t} = \frac{F_n e^{i(\omega - \omega_n)t}}{2} \tag{12}$$

where $\Delta\omega_{klmn} = \omega_k - \omega_l + \omega_m - \omega_n$. Note that for the rotating frame approximation to apply we must have $\Delta\omega_{klmn} \ll \omega$, i.e. separation between the relevant modes must be small compared to the frequency of the modes themselves.

Equation (12) is identical to that describing optical frequency combs (Chembo and Yu 2010). Thus modal analysis of the Euler-Bernoulli equation demonstrates a close parallel with optics concerning the governing dynamics of Kerr combs.

## 2.2 Stability analysis for mechanical comb generation

The task ahead is to determine the parameter space necessary for such comb generation. The below stability and comb threshold analysis follows the same logic presented for optical frequency combs (Matsko et al. 2005, Chembo and Yu 2010, Chembo and Menyuk 2013, Godey et al. 2014) while accounting for the specificity of mechanical systems.

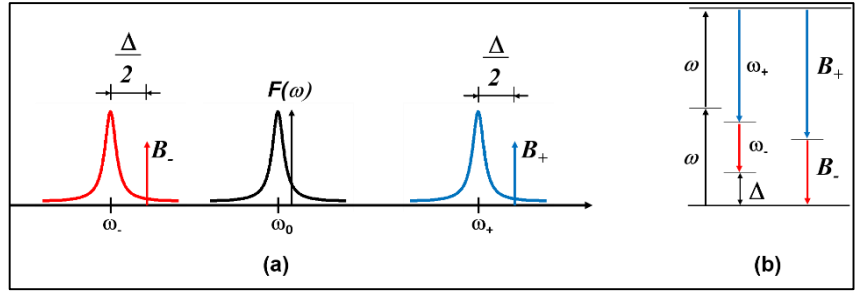

**Fig. 1** Schematic representing the envelope of the driven mode "0", the detuned drive force "$F(\omega)$", the envelopes of the adjacent modes "+" and "-", and the effect of dispersion $\varDelta$ on the generated sidebands $B_{\pm}$ (a). And in (b) schematic representation showing the necessity of introducing the modified rotating frame $B_{\pm}$ to adjust the phonon energy.

It is possible to define the comb generation threshold as the minimum necessary amplitude and detuning for a given driven mode to undergo auto-parametric conversion and generates sidebands that are (nearly) resonant with adjacent modes.

From the above definition, we base our threshold analysis on only three modes, a driven mode (denoted as mode "0"), and two adjacent modes (denoted as "-" and "+" respectively) as shown schematically in Fig. 1a, knowing that by adjacent we mean nearby and not necessarily nearest. We further define:

$$\begin{cases} \Lambda_{00} \cong \Lambda_{++} \cong \Lambda_{--} \equiv \Lambda \\ \Lambda_{-0} = \Lambda_{0-} \cong \Lambda_{0+} = \Lambda_{+0} \equiv \Gamma \\ \Lambda_{+-} = \Lambda_{-+} \equiv \Pi \end{cases} \qquad (13)$$

Since only mode "0" is driven, the right hand side of equation (12) is equated to zero for modes "+" and "-".

The summation corresponding to the nonlinear terms can be further simplified by dropping all second order terms in $A_+$ or $A_-$, equation (12) now reads for the "+" and "-" modes respectively:

$$\begin{cases} \dot{A}_+ = \frac{-\gamma}{2}A_+ + i\frac{6}{8\omega_+}(2(\Lambda^2+2\Gamma^2)|A_0|^2 A_+ + (\Pi\Lambda+2\Gamma^2)A_0 A_-^* A_0 e^{i\Delta t}) \\ \dot{A}_- = \frac{-\gamma}{2}A_- + i\frac{6}{8\omega_-}(2(\Lambda^2+2\Gamma^2)|A_0|^2 A_- + (\Pi\Lambda+2\Gamma^2)A_0 A_+^* A_0 e^{i\Delta t}) \end{cases} \quad (14)$$

where $\Delta = 2\omega - \omega_+ - \omega_- = 2\omega_0 + 2\delta\omega_0 - \omega_+ - \omega_- = \Delta_0 + 2\delta\omega_0$, and $\Delta_0$ is a measure of dispersion, i.e. non-uniform spacing of adjacent modes, whereas $\delta\omega_0$ denotes detuning.

In the above equation, the modal damping coefficients are assumed to take the same value for all modes, a very reasonable assumption for nearby mechanical modes. Since $A_0$ is directly driven, and below the auto-parametric threshold the amplitudes $A_+$ and $A_-$ are negligibly small, thus $A_0$ can simply be determined by solving the classical Duffing equation (Cleland 2003):

$$-\delta\omega_0^2 B_0 + \frac{i}{2}\omega_0\gamma B_0 + \frac{3}{8}6\Lambda^2|B_0|^2 B_0 = \frac{F_0}{2} \quad (15)$$

where $B_0 = A_0 e^{i(\omega_0-\omega)t}$, and $|B_0| = |A_0|$ is a rotating frame transformation used to remove the time dependence in the right hand side of equation (12).

At this point one more rotating reference frame transformation is undertaken whereby the complex amplitudes $B_\pm$ are defined as $B_\pm = A_\pm e^{i\frac{\Delta_t}{2}t}$. Thus equation (14) now reduces to:

$$\begin{cases} \dot{B}_+ = -i\frac{\Delta}{2}B_+ - \frac{\gamma}{2}B_+ + i\frac{6}{8\omega_+}(2(\Lambda^2+2\Gamma^2)|B_0|^2 B_+ + (\Pi\Lambda+2\Gamma^2)B_0 B_-^* B_0) \\ \dot{B}_-^* = i\frac{\Delta}{2}B_-^* - \frac{\gamma}{2}B_-^* - i\frac{6}{8\omega_-}(2(\Lambda^2+2\Gamma^2)|B_0|^2 B_-^* + (\Pi\Lambda+2\Gamma^2)B_0^* B_+ B_0^*) \end{cases} \quad (16)$$

The introduction of the $B_\pm$ coordinates has for objective to eliminate any frequency (energy) mismatch between the pump mode (phonons) and the sideband modes (phonons) as shown schematically in Fig. 1.b. Equation (16) represents an autonomous system that is fully defined by the system parameters, the Jacobian matrix of which is:

$$J = \begin{bmatrix} -i\frac{\Delta}{2} - \frac{\gamma}{2} + \frac{i6}{4\omega_+}(\Lambda^2+2\Gamma^2)|B_0|^2 & \frac{i6}{8\omega_+}(\Pi\Lambda+2\Gamma^2)B_0^2 \\ \frac{-i6}{8\omega_-}(\Pi\Lambda+2\Gamma^2)B_0^{*2} & i\frac{\Delta}{2} - \frac{\gamma}{2} - \frac{i6}{4\omega_-}(\Lambda^2+2\Gamma^2)|B_0|^2 \end{bmatrix} \quad (17)$$

The mechanical Kerr comb generation requires that the eigenvalues ($\lambda$) of the Jacobian matrix $\boldsymbol{J}$ have a positive real part, i.e. $Re(\lambda) > 0$. The threshold of instability, i.e. comb generation, is $Re(\lambda) = 0$ which can be explicitly written as (note that in the denominator we apply the approximation $\omega_0 \cong \omega_+ \cong \omega_-$):

$$\left[-\left(-\Delta + \frac{3}{\omega_0}(\Lambda^2 + 2\Gamma^2)|B_0|^2\right)^2 + \frac{9}{4\omega_0^2}(\Lambda\Pi + 2\Gamma^2)^2|B_0|^4\right] > \gamma^2 \qquad (18)$$

### 2.3  Exploring parameter space

To determine where in the parameter space it is most promising to look for mechanical frequency combs, it is first necessary to determine the main trends of how the nonlinear terms $\Lambda$, $\Gamma$, and $\Pi$ scale with modal number and mode spacing.

Since for the above rotating frame analysis to be valid we need $\omega \gg \Delta$, we consider only large mode numbers, particularly even modes from 50 to 150, so that the spacing between consecutive modes would be much smaller than the mode frequencies.

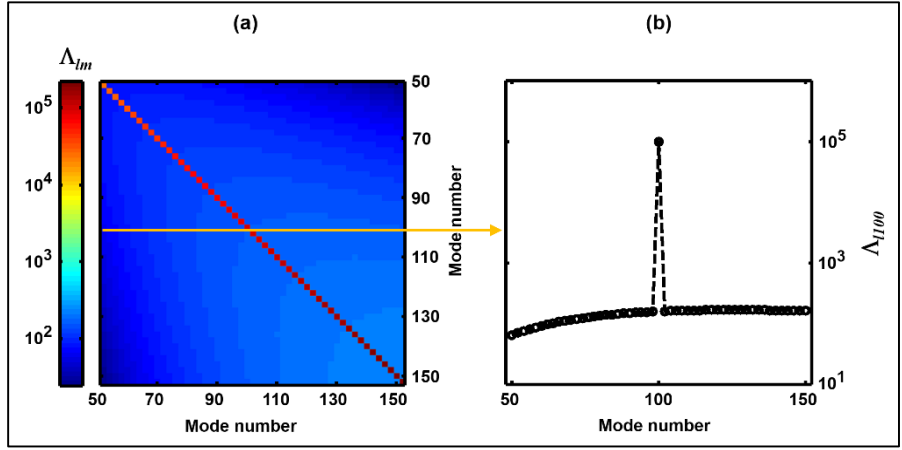

**Fig. 2** Calculation of $\Lambda_{lm}$ showing a strong self-interaction term compared to cross-interaction for modes ranging from 50 to 100 (a). A plot of $\Lambda_{l100}$ for mode number 100 showing 3 orders of magnitude difference between self- and cross- Kerr terms (b).

The coupling terms are calculated by numerically integrating the spatial derivatives of the mode shapes, as explained in section 2.1, the interaction matrix of which is shown in Fig. 2a. Note that the diagonal terms of the figure represent $\Lambda$, whereas off-diagonal terms represent $\Gamma$. It is clear that self- terms are much larger than cross-terms, furthermore cross-interaction terms change very slowly with mode number as can be seen for example from the line plot for the 100th mode shown in Fig. 2b.

Based on the result of these simulations we can approximate $\Gamma \sim \Pi$, and knowing that $\Lambda >> \Gamma$, therefore it is possible to simplify expression (18) into:

$$\left[ -\left( -\Delta + \frac{3}{\omega_0} \Lambda^2 |B_0|^2 \right)^2 + \frac{9}{4\omega_0^2} (\Lambda \Pi)^2 |B_0|^4 - \gamma^2 \right] > 0 \qquad (19)$$

Equation (19) indicates that for comb generation to take place, the nonlinear (Duffing) detuning has to cancel the dispersion, i.e. $\omega_0 \Delta = 3\Lambda^2 |B_0|^2$, and at the same time the cross-Kerr term has to overcome damping, i.e. $3\Lambda\Pi|B_0|^2 = 2\gamma^2$. The threshold values for the amplitude of the pump mode $A_0$ necessary to induce FWM is determined by setting equation (19) to zero, resulting in the following quadratic equation:

$$\frac{9}{\omega_0^2} \Lambda^2 \left( \frac{\Pi^2}{4} - \Lambda^2 \right) |B_0|^4 + \frac{6}{\omega_0} \Delta \Lambda^2 |B_0|^2 - (\Delta^2 + \gamma^2) = 0 \qquad (20)$$

Thus the necessary amplitude for comb generation can be given as function of the system parameters as:

$$|B_0|^2 = \frac{2\omega_0 \left( 2\Delta\Lambda \pm \sqrt{\Delta^2\Pi^2 + \gamma^2\Pi^2 - 4\gamma^2\Lambda^2} \right)}{3\Lambda(4\Lambda^2 - \Pi^2)} \qquad (21)$$

The boundary of the comb generation area is limited by $\gamma = 0$, thus the pump amplitude will need to always be contained between:

$$|B_0|^2 = \frac{4\omega_0 \Delta (\Lambda \pm \Pi/2)}{3\Lambda(4\Lambda^2 - \Pi^2)} \cong \frac{\omega_0 \Delta}{3\Lambda^2} \left( 1 \pm \frac{\Pi}{2\Lambda} \right) \qquad (22)$$

Because $\Pi/\Lambda \approx 10^{-3}$, equation (22) imposes stringent conditions on the pump amplitude, since it needs to be within a very narrow range given by $\sim \frac{\omega_0 \Delta}{3\Lambda^2} (10^{-3})$.

Furthermore, the necessary oscillation amplitude of the pump mode should not fall within the unstable boundaries of the Duffing equation calculated from (15). Now we express the boundaries of the region of instability of equation (15) as:

$$|B_0|^2 = \frac{8\delta\omega_0^2}{27\Lambda^2} \left( 1 \pm \frac{1}{2} \right) \qquad (23)$$

And rewrite equation (22) as:

$$|B_0|^2 = \frac{\omega_0 \Delta_0 + 2\delta\omega_0^2}{3\Lambda^2} \left( 1 \pm \frac{\Pi}{2\Lambda} \right) \qquad (24)$$

Equations (23) and (24) define the necessary amplitude-detuning space for comb generation area and for the unstable amplitude solution, respectively. Several important distinctions between the optics and mechanics cases are readily visible. For one, in the optics case if $\Delta_0 = 0$ then comb generation does not take place because the slope of the comb generation area is identical to that of the Duffing unstable solution (Chembo and Yu 2010). This is not the case in mechanics, since for $\Pi/\Lambda << 1$ the slope of the comb generation space is quite different than that of the Duffing unstable regime.

A further important distinction between optical and mechanical frequency combs, is that for the latter the calculated Duffing nonlinearity always takes on positive values, i.e. $\Lambda^2 > 0$. This does not change the anomalous dispersion case, i.e. $\Delta_0 < 0$, being the more promising case to achieve frequency comb generation as can be seen in Fig. 3.

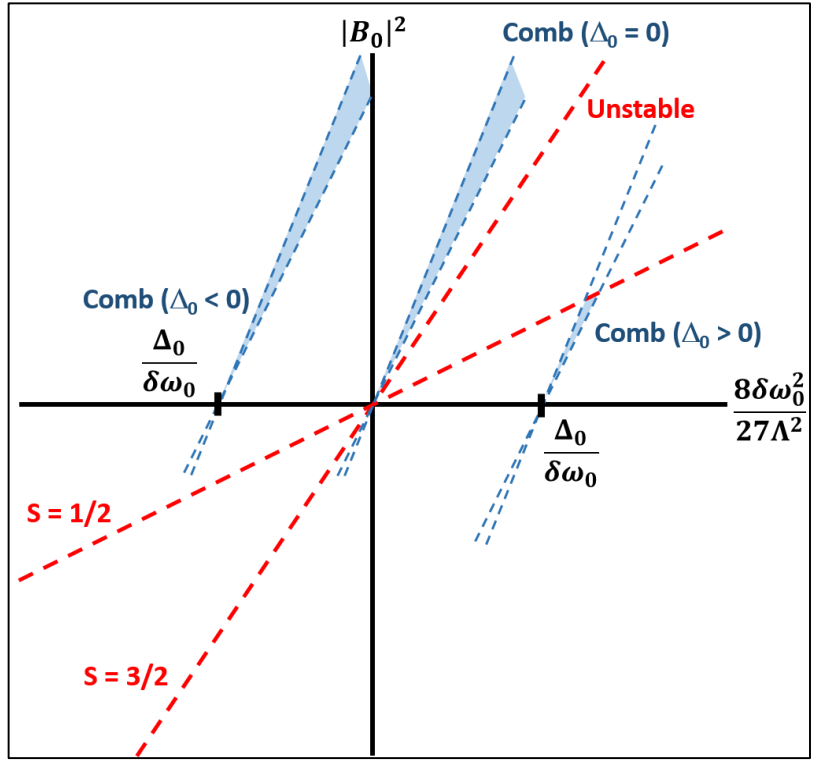

Fig. 3 Comb generation zone in amplitude-detuning parameter space shown in light blue shading, the border of the unstable amplitude solution area is shown with the dashed red lines. Unlike optics, comb generation can still take place if no dispersion exists, i.e. $\Delta_0 = 0$, but requires frequency detuning, i.e. $\delta\omega_0 \neq 0$. Comb generation also takes place if dispersion is introduced, i.e. $\Delta_0 \neq 0$, but the situation with $\Delta_0 < 0$ is more promising. $\Pi/\Lambda$ is taken to be 0.2 to make the parameter space more visible, in reality $\Pi/\Lambda$ is smaller than that.

## 2.4 Effect of internal stresses on comb generation

When fabricating micro- and nanomechanical devices, a great deal of internal structural stresses are encountered that usually originate from the deposition techniques used to produce the fine layering of M/NEMS devices (Laconte et al. 2006, Madou 2011). Such internal stresses can change drastically the behavior of a MEMS device by shifting its resonance frequencies, and changing the mode shapes (Houri et al. 2013).

To account for such possibility we introduce an initial stress term in the dimensionless Euler-Bernoulli equation (3) that now reads:

$$\frac{\partial^4 y}{\partial x^4} + \gamma \frac{\partial y}{\partial t} + \frac{\partial^2 y}{\partial t^2} - \left(12\bar{N} + 6 \int_0^1 \left(\frac{\partial y}{\partial x}\right)^2 dx\right)\frac{\partial^2 y}{\partial x^2} = F \tag{25}$$

where $\bar{N}$ is the normalized internal stress given as: $\bar{N} = \frac{NL^2}{EAh^2}$, where $N$ is the absolute stress, note that for tensile stresses $N$ is positive, and compressive stresses appear as negative $N$. As done previously from hereon we drop the bars for convenience.

Here again the normal modes can be obtained from the homogenous Euler-Bernoulli equation by solving:

$$\frac{d^4\psi(x)}{dx^4} - 12N\frac{d^2\psi(x)}{dx^2} = \omega^2\psi(x) \tag{26}$$

Whose mode shapes are expressed as (Blevins 1979, Tilmans et al. 1992):

$$\psi_n(x) = \cos(\lambda_n x) - \cosh(\mu_n x)$$
$$+ \frac{\cos(\lambda_n) - \cosh(\mu_n)}{-\sin(\lambda_n) + \frac{\lambda_n}{\mu_n}\sinh(\mu_n)}(\sin(\lambda_n x) - \sinh(\mu_n x))$$

$$\tag{27}$$

Where:

$$\begin{cases} \mu_n = \left[6N + \sqrt{37N^2 + \omega_n^2}\right]^{1/2} \\ \lambda_n = \left[-6N + \sqrt{37N^2 + \omega_n^2}\right]^{1/2} \\ \cos(\lambda_n)\cosh(\mu_n) - 0.5\left(\frac{\mu_n}{\lambda_n} - \frac{\lambda_n}{\mu_n}\right)\sinh(\mu_n)\sin(\lambda_n) - 1 = 0 \end{cases} \tag{28}$$

Whereas the modal analysis procedure itself is not altered significantly by the introduction of the initial tension term, the change to the modal shapes has a considerable impact on the cross-Kerr coefficients.

This impact can be conceptually understood by considering that beam structures under very high tensile stresses can be described as strings (Verbridge et al. 2006) where the fourth order spatial derivative is dropped, and for such second order differential equation the cross-Kerr terms are zero. Thus the more the structure is subjected to tensile stresses the more the mode shapes resembles pure sinusoids, and the smaller the cross-Kerr terms are. The impact of internal stress is demonstrated in Fig .4 where the terms $\Lambda_{33}$, and $\Pi_{35}$ are shown as a function of normalized tension.

Thus, internal stresses (tension in particular) suppress the formation of frequency combs, and inhomogeneous stresses most likely make it impossible.

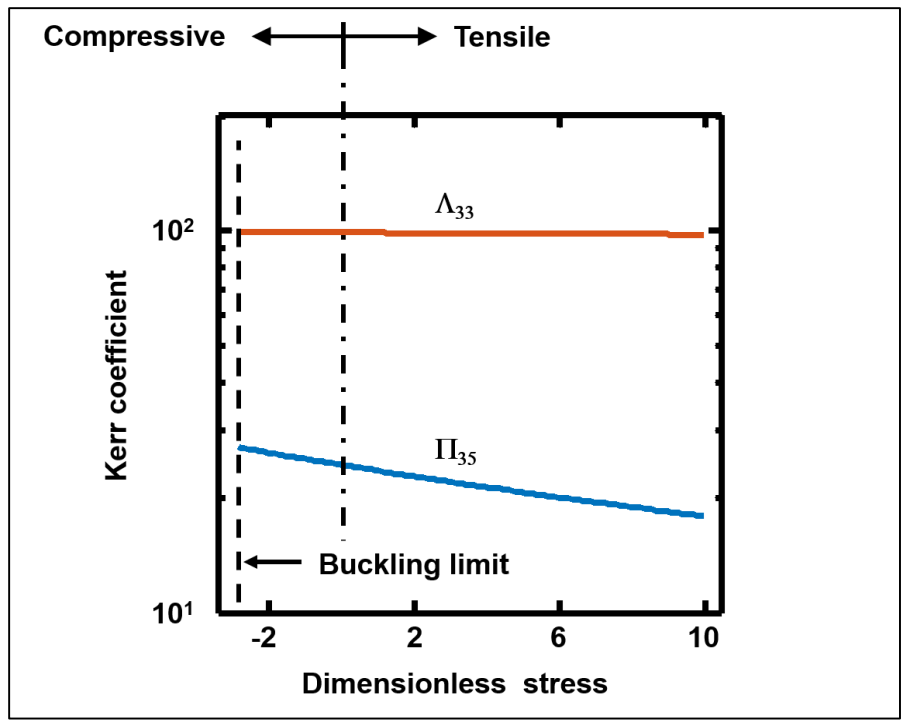

**Fig. 4 Effect of internal stress on the coupling parameters $\Pi_{35}$ and $\Lambda_{33}$, showing that the cross-Kerr term drops quickly as the normalized stress is increased. Note that for the small mode numbers used here (modes 3 and 5) the ratio $\Pi/\Lambda \sim 10^{-1}$ which is 2 orders of magnitude larger than for large mode numbers.**

## 3    Towards experimental realization

The modal analysis performed so far has relied on the approximation that the frequency spacing between the modes is much smaller than the frequency of the modes themselves. If one desires such a condition to be true for simple Euler-Bernoulli clamped-clamped beams, then the range of validity is restricted to high mode numbers.

Nevertheless, high mode numbers come with their own problems, most notably the $\Pi/\Lambda \sim 10^{-3}$ ratio calculated in section 2.3 for mode numbers from 50 to 150, are particularly small, making the area for comb generation within parameter space miniscule and difficult to achieve experimentally. Furthermore, very large mode numbers are experimentally difficult to excite in an efficient manner.

This leads to a seemingly contradictory conclusion where to obtain a large $\Pi/\Lambda$ ratio a low mode number is required whereas to maintain the approximation of $\Delta \ll \omega$ a high mode number is required.

A promising workaround consists in using mechanical Fabry-Perot like structures (Hatanaka et al. 2014, Hatanaka et al. 2015, Cha and Daraio 2018, Kurosu et al. 2018, Hatanaka et al. 2019). These structures are in fact rectangular suspended plates, as shown schematically in Fig. 5a, where one of the plate dimensions is much smaller than the other, say *length >> width*. Despite the plate equation being more cumbersome, the nonlinear vibrations of a plate can be reduced to an equation similar to equation (10) that governs the multimode nonlinear response of Euler-Bernoulli beams (Crawford and Atluri 1975, Nayfeh and Mook 1979, Chia 1980, Sathyamoorthy 1997, Nayfeh 2000).

Unlike beams, plate geometry induces a cutoff frequency set by the smaller of the two geometrical dimensions. The larger the length-to-width ratio of the mechanical cavity is, the higher the cavity cutoff frequency and the closer the modes are to each other. This is shown in the experimentally measured frequency response of a piezoelectric GaAs mechanical Fabry-Perot cavity with a width of 20 μm and a length of 1 mm, Fig. 5b. Thus for a Fabry-Perot type mechanical cavity it is possible to satisfy both the condition $\Delta \ll \omega$ and the condition of a small modal number simultaneously. In addition nonlinear response was demonstrated for such structures (Hatanaka et al. 2017), in the form of non-degenerate FWM under the effect of dual tone drive.

Note that the microfabrication process necessitates to have periodically located etch holes in the structures. These are simply small circular openings that allow the chemical etching of the underlying material and thus the release and suspension of the plate structure. The presence of such periodic lattice of holes induces bandgaps in the frequency response of the cavity. The effects of the latter is to change the dispersion relation as the bandgap is approached, whereas the system exhibits anomalous dispersion after the cutoff, i.e. $\Delta_0 < 0$, the sign of dispersion changes as we approach the bandgap. An effect that is visible in Fig. 5c for the experimentally obtained values of $\Delta_0$.

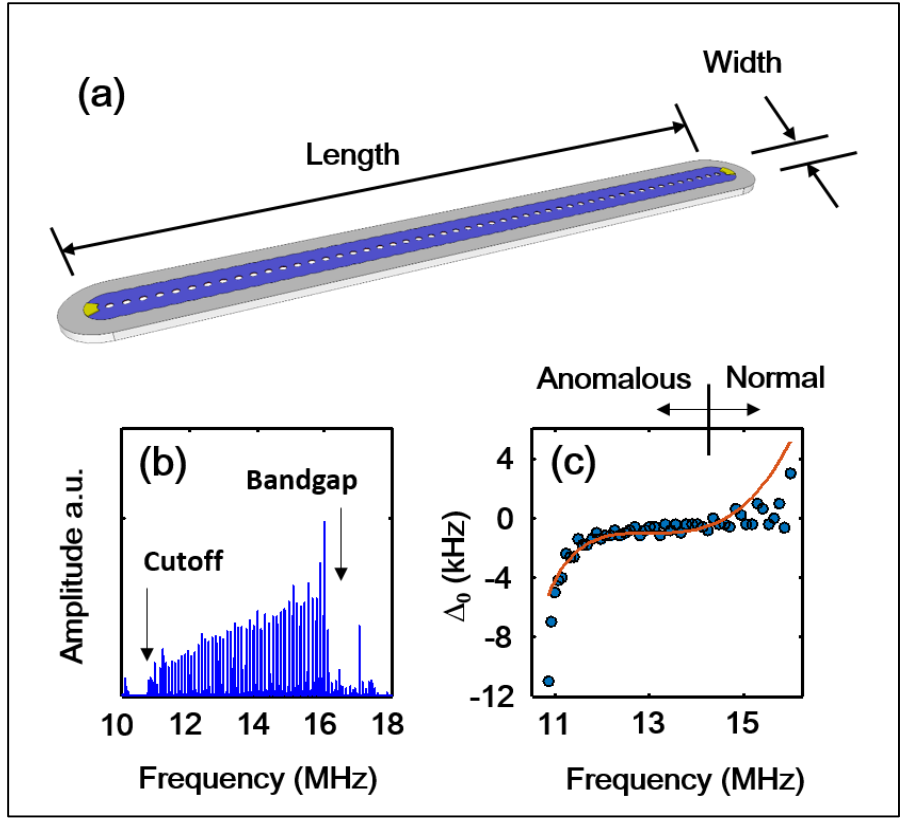

**Fig. 5** schematic representation of a mechanical Fabry-Perot cavity with *length >> width,* showing the suspended portion (in blue) and the periodically located etch holes (a). The measured frequency response of a mechanical cavity having width = 20 μm, and length = 1 mm (b) clearly showing the cutoff and bandgap. From the resonance frequencies shown in (b) it is possible to extract the dispersion $\Delta_0$ of the structure, which despite the scatter shows a negative value, corresponding to anomalous dispersion, after the cutoff and positive values, corresponding to normal dispersion, nearing the bandgap (c).

## 5    Conclusion

In summary this work demonstrated that starting from a generic description of a beamlike mechanical MEMS structure, it is possible to derive through a modal analysis approach a governing equation that is qualitatively identical to that governing optical frequency combs. Thereafter the possibility of generating mechanical frequency combs was investigated in depth to determine the necessary experimental conditions. The analysis performed led to a seemingly contradictory set of conditions where on the one hand it required a large modal number for the approximations made to hold, while on the other hand a large modal number gives an impractically small area for comb generation.

A workaround was suggested to employ rectangular plate structures with a high width to length ratio, these structures preserve the dynamics to a large extent while alleviating the conditions on frequency separation. In addition the presence of a periodic lattice in the suspended membrane induces bandgaps that also provide means for dispersion engineering.

Despite these early encouraging results, their remain several open question regarding the scaling of nonlinear Kerr coefficients in a plate-like structure, and the experimental difficulties in exciting sufficiently large amplitudes for multimodal mechanical comb generation to take place.

## Acknowledgements

This work is partly supported by a MEXT Grant-in-Aid for Scientific Research on Innovative Areas "Science of hybrid quantum systems" (Grant No. JP15H05869 and JP15K21727).